\documentclass[10pt, conference, compsocconf]{IEEEtran}
\IEEEoverridecommandlockouts
\usepackage{cite}
\usepackage{color}
\usepackage{graphicx}
\usepackage{amssymb,wasysym}
\usepackage[vlined,linesnumbered,ruled,norelsize]{algorithm2e}

\newtheorem{lemma}{\textbf{Lemma}}
\newtheorem{theorem}{\textbf{Theorem}}
\newtheorem{prop}{\textbf{Proposition}}

\begin{document}

\title{MEGCOM: Min-Energy Group COMmunication in Multi-hop Wireless Networks}

\author{\IEEEauthorblockN{
Kai Han \IEEEauthorrefmark{1}\IEEEauthorrefmark{2},
Liu Xiang \IEEEauthorrefmark{2},
Jun Luo\IEEEauthorrefmark{2}, and
Yang Liu \IEEEauthorrefmark{3}
}
\IEEEauthorblockA{\IEEEauthorrefmark{1}School of Computer Science, Zhongyuan University of Technology, China\\}
\IEEEauthorblockA{\IEEEauthorrefmark{2}School of Computer Engineering, Nanyang Technological University, Singapore\\}
\IEEEauthorblockA{\IEEEauthorrefmark{3}School of Information Sciences and Engineering, Henan University of Technology, China\\
\{hankai, xi0001iu, junluo\}@ntu.edu.sg, enjoyang@gmail.com}
}

\maketitle

\begin{abstract}
  Given the increasing demand from wireless applications, designing energy-efficient group communication protocols is of great importance to multi-hop wireless networks. A group communication session involves a set of member nodes, each of them needs to send a certain number of data packets to all other members. In this paper, we consider the problem of building a shared multicast tree spanning the member nodes such that the total energy consumption of a group communication session using the shared multicast tree is minimized. Since this problem was proven as NP-complete, we propose, under our Min-Energy Group COMmunication (MEGCOM) framework, three distributed approximation algorithms with provable approximation ratios. When the transmission power of each wireless node is fixed, our first two algorithms have the approximation ratios of $\mathcal{O}\left(\ln(\Delta + 1)\right)$ and $\mathcal{O}(1)$, respectively, where $\Delta$ is the maximum node degree in the network. When the transmission power of each wireless node is adjustable, our third algorithm again delivers a constant approximation ratio. We also use extensive simulations to verify the practical performance of our algorithms.
\end{abstract}

\IEEEpeerreviewmaketitle

\section{Introduction} \label{sec:intro}
  Group communication (or all-to-all multicasting) is a very important primitive for distributed systems, as a large body of applications including, among others, social networking, online meeting, network gaming, resource sharing, and data management are heavily relying on its service \cite{Powell-CACM96,Chockler-CS01}. In a traditional setting (e.g., the Internet), investigations on group communication focus on the system reliability in the face of network failures or group member changes \cite{Chockler-CS01}. As the recent booming of multi-hop wireless networks for various purposes (e.g., wireless mesh networks to allow Internet access in remote areas or wireless sensor networks for habitat monitoring) further demand group communication to work on top of such networks (e.g., \cite{YuC-MSWiM09}), group communication is facing a new challenge: energy constraints for wireless nodes. As it is generally believed that the continuous development of the wireless technology will result in more and more wireless applications making use of group communication primitive (e.g., mobile social networking and gaming), designing energy-efficient group communication protocols has become imperative.

  One naive way for designing an energy-efficient group communication protocol is to employ the existing one-to-many multicast protocols. In other words, one may construct an energy-efficient one-to-many multicast tree for each group member. While this approach may leverage on the abundant research results on building minimum-energy (one-to-many) multicast tree in wireless networks (e.g., \cite{Wieselthier2000,Wan2004,Liang2006,Li2007}), it is not exactly practical, as it would lead to a huge computation and communication overhead in maintaining all these multicast trees. Therefore, a more practical way is to construct one tree spanning the group members as a shared multicast tree.

  Surprisingly, although the minimum energy one-to-many multicasting problem has been studied extensively, there is little work on min-energy group communication in multi-hop wireless networks. Similar to the min-energy one-to-many multicasting problem, the min-energy shared tree multicasting problem is also NP-hard \cite{Liang2009}. Moreover, the problem becomes more challenging because an optimal tree also depends on how many packets each group member is about to multicast. Liang \textit{et al.}\ \cite{Liang2009} attempted to tackle this problem by proposing several approximation algorithms. However, the resulting approximation ratios are far from satisfactory. For example, their approximation ratios are in the same order of $|M|$, where $M$ is set of group members.

  In this paper, we study the Min-Energy Group COMmunication (MEGCOM) framework in multi-hop wireless networks. In particular, we seek to minimize the total energy consumption of a group communication session relying on a shared multicast tree. As the induced Minimum-Energy All-to-All Multicasting (MEAAM) problem is NP-hard, we propose three distributed approximation algorithms with guaranteed approximation ratios for MEAAM, under both cases of fixed transmission power and of adjustable transmission power. Our algorithms significantly improve the best known results. Specifically:
  \begin{enumerate}
    \item When the transmission power of each wireless node is fixed, our first algorithm has an approximation ratio of $4\ln(\Delta+1)+7$, where $\Delta$ is the maximum node degree in the network. 
    \item When the transmission power of each wireless node is fixed, our second algorithm delivers a 13-approximation to MEAAM. 
    \item When the transmission power of each wireless node is adjustable, we prove the existence of a constant approximation algorithm for MEAAM, and we also show that a straightforward algorithm leads to a constant approximation ratio of 145. 
  \end{enumerate}

  In the remaining of the paper, we first briefly review the literature in Sec.~\ref{sec:rlw}. After formally defining the models and problem in Sec.~\ref{sec:model}, we present and analyze our three algorithms in Sec.~\ref{sec:logftxp}, \ref{sec:conftxp}, and \ref{sec:conatxp} respectively. We also perform extensive simulations of our algorithm in Sec.~\ref{sec:sim}, and we finally conclude our paper in Sec.~\ref{sec:con}.

\section{Related Work} \label{sec:rlw}
  There is a large body of literature on group communication and multicasting, but, given the space limitation, we have to confine our discussions to those aiming at minimizing the energy consumption, but to leave out non-tree-based appraoches such as \cite{LeeSM-MONET02,TavliH-TC11}.

  The minimum-energy one-to-many multicasting problem has been studied in \cite{Wieselthier2000,Wan2004,Liang2006,Li2007}. Wieselthier \textit{et al.}\ \cite{Wieselthier2000} consider a scenario where each node can adjust its transmission power continuously, and proposed several greedy heuristics for the minimum-power broadcast/multicast routing problems. Wan \textit{et al.}\ \cite{Wan2004} prove that the heuristics proposed by \cite{Wieselthier2000} have linear approximation ratios, and provide several approximation algorithms with constant approximation ratios for the minimum-energy multicasting problem based on the approximate minimum Steiner tree algorithm. Liang \cite{Liang2006} consider a scenario in which each wireless node can adjust its transmission power in a discrete fashion, and the communication links are symmetric. They propose a centralized approximation algorithm with performance ratio $\mathcal{O}(\ln K)$ for building a minimum-energy multicasting tree, where $K$ is the number of destination nodes in a one-to-many multicast session. Li \textit{et al.}\ \cite{Li2007} consider a case where all nodes have a fixed transmission power and the communication links are asymmetric. They convert the minimum-energy multicasting problem to an instance of the Directed Steiner Tree problem, and presented several heuristics.

  To the best of our knowledge, the only work that studies the minimum-energy group communication problem is \cite{Liang2009}. In \cite{Liang2009}, Liang \textit{et al.}\ propose to build a shared multicast tree such that the total energy consumption of realizing a group communication session using the shared tree is minimized. They prove that finding such a shared tree is a NP-complete problem, and used the approximate Steiner tree algorithm proposed by \cite{Kou1981} to solve the problem. When the transmission power of each node is fixed, they prove that the approximate minimum Steiner tree has an approximation ratio of $2(|M|+1)$, where $M$ is the set of group members. When the transmission power of each node is adjustable, they prove that the approximate Steiner tree has an approximation ratio of $\Omega(8|M|)$. They also proposed a distributed approximation algorithm with an approximation ratio of $\Omega(4|M|^2)$.

\section{Preliminaries and Problem Definition} \label{sec:model}
  A  multi-hop wireless network is modeled by an undirected graph $G=(V,E)$, where $V$ is the set of wireless nodes in the network and $E$ is the set of wireless links. Each node in $v \in V$ has a unique identifer $v.\mathit{id}$. The nodes in $V$ are
  all equipped with an omni-directional antenna. We assume that the transmission (tx) power of each node can be either identically fixed or continuously adjustable. When the tx power is fixed, we use $\varepsilon_s$ to denote the energy consumption of transmitting a data packet by a node. When the tx power is adjustable, the energy required by any node $u$ to transmit a data packet to another node $v$ can be determined by the Euclidean distance between $u$ and $v$. Following a very common formula, we define such energy consumption to be $d_{(u,v)}^\alpha$, where $d_{(u,v)}$ is the Euclidean distance between $u$ and $v$ and $\alpha$ is a constant (usually between 2 and 4). In either case, we assume that each link $(u,v)\in E$ is assigned a weight which is the amount of energy required for $u$ to send a data packet to $v$ (or vise versa). We also assume that the receiving (rx) power is always less than the tx power, and we denote by $\varepsilon_r$ the energy consumption of receiving a data packet by a node.

  In a group communication session, there exists a set of \textit{group members} $M\subseteq V$, and each node $u \in M$ needs to send $p(u)$ data packets to all other nodes in $M\backslash \{u\}$. We denote by $k$ the sum of the numbers of data packets originated from the group members, i.e., $k = \sum\nolimits_{v \in M} {p(v)}$. As we explained in Sec.~\ref{sec:intro}, instead of building $|M|$ multicast trees originated from each node in $M$, building a shared multicast tree spanning the nodes in $M$ to support the group communication session is more convenient in realistic settings. Therefore, in order to minimize the total energy consumption of the group communication session, we actually aim at solving the following Minimum-Energy All-to-All Multicasting (MEAAM) problem:
  \begin{quote}
    Find an \textit{optimal shared multicast tree} $T_{\mathit{opt}}$ such that the energy consumption of carrying all $k$ packets over $T_{\mathit{opt}}$ is minimized.
  \end{quote}

  The hardness of this problem is immediate from \cite{Liang2009}:
  \begin{prop} \label{prp:NPC}
    The MEAAM problem is NP-complete.
  \end{prop}

  For convenience of description, we define some other notations here. For any multicast tree $T$ in $G$, we denote by $\mathit{nd}(T)$ the set of nodes in $T$, by $\mathit{lv}(T)$ the set of degree-one nodes in $T$, by $\mathit{in}(T)$ the set of internal nodes (nodes with degree more than one) in $T$, and by $\Psi(T)$ the total energy consumption of realizing a group communication session using $T$. For a node $u \in V$, we denote the set of neighboring nodes of $u$ in $G$ by $\mathit{nb}(u,G)$, and let $\mathit{nb}^+(u,G) = \mathit{nb}(u,G) \cup \{u\}$ and $\mathit{nb2}(u,G) = \bigcup_{v \in \mathit{nb}(u,G)} \mathit{nb}^+(v,G)$, where the latter is the two-hop neighbors of $u$. In the following, we present three distributed algorithms to construct approximate trees for $T_{\mathit{opt}}$, they respectively achieve a logarithmic approximation for fixed tx power, a constant approximation for fixed tx power, and a constant approximation for adjustable tx power.

\section{MEGCOM-LFP: Logarithmic Approximation for MEAAM with Fixed tx Power} \label{sec:logftxp}
  In this section, we propose a distributed approximation algorithm for MEAAM with fixed tx power. A brief introduction on the idea of the algorithm comes first, followed by detailed algorithm descriptions and a performance analysis. We also discuss how to maintain the shared group communication tree in the face of member joining and leaving.

  \subsection{Algorithm Principles} \label{sec:principle}
\begin{figure}[htbp]
      \centerline{\includegraphics[width=\columnwidth]{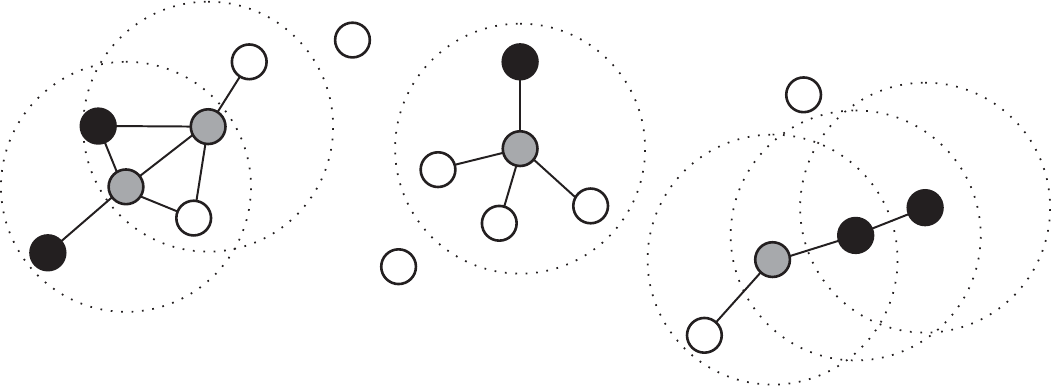}}
      \caption{Examples for the buddy set selection. The black nodes are group members in $M$, the grey nodes are the non-member nodes chosen for $B$. Here $B$ contains every node having a concentric tx circle surrounding it; the solid lines indicate edges (or links) implied by those tx circles.}
      \label{fig:cover}
    \end{figure}

    Unlike the common wisdom that directly uses a Steiner tree to approximate $T_{\mathit{opt}}$, our algorithm has two stages (with the first one having two sub-stages) to construct a $T_A$ that approximates $T_\mathit{opt}$:
    \begin{list}{\labelitemi}{\itemindent=-1.5em}{\listparindent=2em}{\labelwidth=3em}
      \item[S1-a:] Identify a \textit{buddy set} $B \subseteq V$ such that, for each $v \in B$, $\mathit{nb}(v, G) \cap M \not= \emptyset$.
      \item[S1-b:] Find a \textit{guardian set} $C \subseteq B$ for $M$, such that $M\subseteq \bigcup_{v\in C}\mathit{nb}^+(v,G)$.
      %
      \item[S2~~:] Construct an approximate Steiner tree $T_A$ to span the nodes in $C$.
    \end{list}
    As a result, $T_A$ is the union of the member set $M$, the guardian set $C$ (along with edges it uses), and the Steiner tree that spans $C$. We term each $v \in C$ a \textit{guarding node} and a member $u \in M \wedge u \in \mathit{nb}(v, G)$ the \textit{guarded member} of $v$. The tricky part is that we identify $C$ by only searching among $B$ that contains nodes having at least one neighbor in $M$, which excludes nodes in $M$ that has no neighbor in $M$. As shown in Figure~\ref{fig:cover},
    all non-member nodes having at least one member in its neighborhood belong to $B$, whereas a member node belongs to $B$ only if it has at least one other member in its neighborhood.

\begin{figure}[htbp]
      \centerline{\includegraphics[width=\columnwidth]{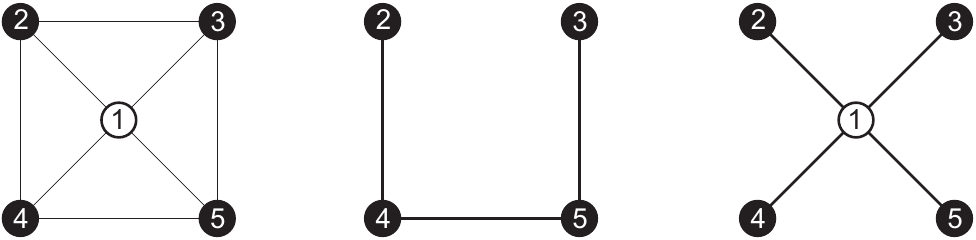}}
      \parbox{.32\columnwidth}{\center\scriptsize(a)~~~~~~~}
      \parbox{.32\columnwidth}{\center\scriptsize(b)}
      \parbox{.32\columnwidth}{\center~~~~~~~\scriptsize(c)} \\ \\
      \centerline{\includegraphics[width=.85\columnwidth]{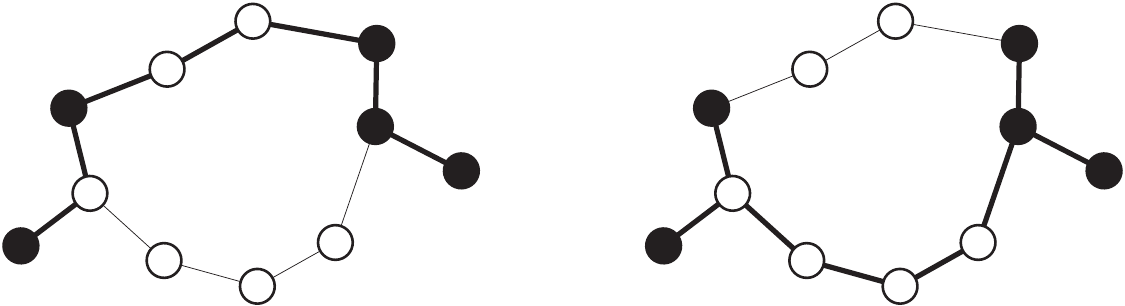}}
      \parbox{.49\columnwidth}{\center\scriptsize(d)}
      \parbox{.49\columnwidth}{\center\scriptsize(e)}
      \caption{Comparing our algorithm with Steiner tree based on two instances of MEAAM. In the first example (a)--(c), the group members is $\{2,3,4,5\}$. The numbers of data packets originated from nodes $2$, $3$, $4$, and $5$ are 100, 100, 1, and 1, respectively. The tx power of each node is 10. The reception power at each node is 1. The optimal Steiner tree in (b) leads to a total energy consumption of 6646, while our multicast tree (optimal for this case) in (c) gives 4848. In the second example (d)--(e), a Steiner-based heuristic (d) can result in 6 nodes in $\mathit{in}(T_A)$, whereas our multicast tree (e) only has 5 such nodes.}
      \label{fig:example}
    \end{figure}

    The motivation behind our algorithm is that a good approximation $T_A$ should have more nodes in $\mathit{lv}(T_A)$ and less nodes in $\mathit{in}(T_A)$, as a node in $\mathit{in}(T_A)$ consume tx power for all packets and rx power for those not originated from it, whereas a node in $\mathit{lv}(T_A)$ consumes only tx power for packets originated from it and otherwise consume only rx power. We use two examples in Figure~\ref{fig:example} to illustrate the advantage of trees constructed by our algorithm and straightforward Steiner trees that span $M$.
    Therefore, we argue that using a Steiner-based heuristic does not give a proper approximation to $T_{\mathit{opt}}$. By far, the best-known centralized approximation algorithm for MEAAM applies a Steiner-based heuristic \cite{Liang2009}, it only has an approximation ratio of $2(|M|+1)$. The performance analysis of our algorithm in Sec.~\ref{sec:perf1} shows that our algorithm achieves a much better approximation ratio.

  \subsection{Algorithm Details}
\begin{figure}[htbp]
      \centerline{\includegraphics[width=.8\columnwidth]{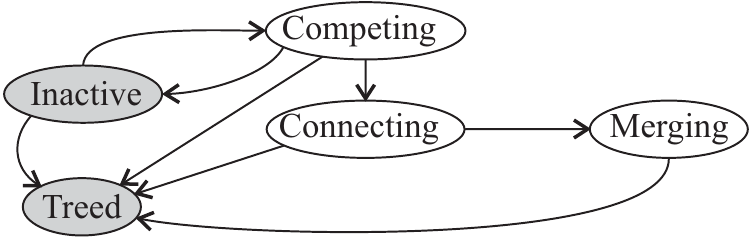}}
      \caption{The finite state machine for each node.}
      \label{fig:state}
    \end{figure}

    As explained in Sec.~\ref{sec:principle}, our algorithm has two stages. In fact, each node can be in different states in each stage, so we let each node $v \in V$ to maintain a variable $v.\mathit{state}$ in order to keep track of the algorithm progress. Initially, every node starts with a state \textbf{Inactive}, and the algorithm terminates when all members in $M$ are \textbf{Treed} (i.e., join the shared group communication tree) and all \textbf{Treed} non-members are spanned by shortest paths to form an approximate Steiner tree. We show the finite state machine for each node in Figure~\ref{fig:state}.
    The state \textbf{Competing} is involved in the guardian set identification (S1), while the two states \textbf{Connecting} and \textbf{Merging} are for approximate Steiner tree construction (S2).

    We first present the pseudocodes of our algorithm by \textbf{Algorithm~\ref{alg:fixedplog}}, then we explain the algorithm execution in details.
    \begin{algorithm}[h!]
      \KwIn{The network $G=(V,E)$ and the member set $M$. For each node $v \in V$, $v.\mathit{state} = \mathbf{Inactive}$}
      \KwOut{An approximate multicast tree $T_A$: all the \textbf{Treed} nodes and all the marked paths/edges}
      \hrule
      \textbf{task}~\textsc{Scan}$()$  \hfill /* \textsf{\small executed periodically} */ \label{ln:scanb} \\
      \uIf {$v.\mathit{state} = \mathbf{Inactive} \wedge \exists u \in \mathit{nb}(v,G): u \in M$} { \label{ln:coverb}
        \ForAll {$u \in \mathit{nb}(v,G): u \in M$} {
          \If {$u.state = \mathbf{Inactive}~\mathrm{or}~\mathbf{Competing}$} { \label{ln:count}
            $v.S \leftarrow v.S \cup \{u\};~~v.\mathit{ct} \leftarrow v.\mathit{ct} + 1$
          }
        }
        \lIf {$v.\mathit{ct} > 0$} {
          $v.\mathit{state} \leftarrow \mathbf{Competing}$; 
          \textsc{sendCompete}$(v.\mathit{id}, v.\mathit{ct})$ to $\forall u \in \mathit{nb2}(v,G)$ \label{ln:cvcomp}
        }
      }
      \ElseIf {$v.\mathit{state} = \mathbf{Connecting}$} {
        \ForAll {$u \in M$} { \label{ln:pathqb}
          \If {$u.state = \mathbf{Treed}~\mathrm{or}~\mathbf{Connecting}$} {
            $v.S \leftarrow v.S \cup \{u.\mathit{gd}\};~~v.\mathit{ct} \leftarrow v.\mathit{ct} + 1$ \\
            $v.\mathit{Path} \leftarrow v.\mathit{Path} \cup \mathit{SP}(v \leftrightsquigarrow u.\mathit{gd})$ \label{ln:pathqe}
          }
        }
        \If {$v.\mathit{ct} = |M|$} {
          $v.\mathit{prefer} \leftarrow \arg\min_u\{\mathit{SP}(v \leftrightsquigarrow u) \in v.\mathit{Path}\}$ \label{ln:ssp} \\
          \textsc{sendCNTReq}$(v.\mathit{id}, v.\mathit{ld})$ to $v.\mathit{prefer}$ \label{ln:ssps}
        }
      }
      \lElseIf {$v.\mathit{state} = \mathbf{Treed} \wedge v.\mathit{gd} = v$} {
        \textsc{Discover}$()$ \\
      }
      \ElseIf {$v.\mathit{state} = \mathbf{Merging}$} {
        \lIf {$\rightharpoondown$\textsc{Discover}$()$} {$v.\mathit{state} \leftarrow \mathbf{Treed}$} \label{ln:terminate} \\
        \lElse {\textsc{sendCNTReq}$(v.\mathit{id}, v.\mathit{ld})$ to $v.\mathit{prefer}$} \label{ln:scane}
      }
      \hrule
      \textbf{upon}~\textsc{recvCompete}$(u.\mathit{id}, u.\mathit{ct})$ from $\forall u \in \mathit{nb2}(v,G)$ \\
      \uIf {$v.\mathit{ct} = \max_{u \in \mathit{nb2}(v,G)\cup\{v\}}\{u.\mathit{ct}\}$} { \label{ln:winner}
        $v.\mathit{state} \leftarrow \mathbf{Connecting};~~v.\mathit{gd} \leftarrow v;~~v.\mathit{ld} \leftarrow v$ \label{ln:s1tos2} \\
        \textsc{sendDecision}$(v.\mathit{id})$ to $\forall u \in v.S$ \label{ln:notecv}
      }
      \lElse {
        $v.\mathit{state} \leftarrow \mathbf{Inactive}$ \label{ln:confirmcv} \\
      }
      $v.S \leftarrow \emptyset;~~v.\mathit{ct} \leftarrow 0$
      \hrule
      \textbf{upon}~\textsc{recvDecision}$(u.\mathit{id})$ \\
      $v.\mathit{state} \leftarrow \mathbf{Treed};~~v.\mathit{gd} \leftarrow u;$~~mark the edge to $u$
      \hrule
      \textbf{upon}~\textsc{recvCNTReq}$(u.\mathit{id}, u.\mathit{ld})$ \\
      \If {$v.\mathit{prefer} = u$} { \label{ln:match}
        mark the path suggested by the request \\
        \textsc{sendAccept}$(v.\mathit{id}, \min\{v.\mathit{ld},u.\mathit{ld}\})$ to $u$  \\
        \lIf {$v.\mathit{ld} < u.\mathit{ld}$} { \label{ln:sid1}
          $v.\mathit{state} \leftarrow \mathbf{Merging}$ \\
        }
        \lElse {
          $v.\mathit{state} \leftarrow \mathbf{Treed};~~v.\mathit{ld} \leftarrow u.\mathit{ld}$
        }
      }
      %
      \hrule
      \textbf{upon}~\textsc{recvAccept} $(u.\mathit{id}, u.\mathit{ld})$ \\
      mark the accepted path \\
      \lIf {$v.\mathit{ld} < u.\mathit{ld}$} { \label{ln:sid2}
        $v.\mathit{state} \leftarrow \mathbf{Merging}$ \\
      }
      \lElse {
        $v.\mathit{state} \leftarrow \mathbf{Treed};~~v.\mathit{ld} \leftarrow u.\mathit{ld}$
      }
      \caption{Finding an approximate multicast tree $T_A$ for the MEAAM problem} \label{alg:fixedplog}
    \end{algorithm}
    As we do not assume that the distributed tree construction process can start synchronously at all nodes, we set a periodically executed \textit{task} \textsc{Scan}$()$ (lines~\ref{ln:scanb} to \ref{ln:scane}) to drive the process. This task includes two part: lines~\ref{ln:coverb} to \ref{ln:cvcomp} take care of identifying a guardian set $C$ in a \textbf{greedy} manner, while other codes are for Steiner tree. In addition, there are four \textit{procedure}s that respond to different events.

    At the beginning of the guardian set identification stage (S1), every node having at least one member neighbor implicitly joins the \textbf{buddy set} $B$ by changing its state to \textbf{Competing} (line~\ref{ln:cvcomp}), as it has the right to compete for acting as a guarding node. As the guarded (member) sets do not intersect if two buddy nodes are more than two hops way from each other, it is sufficient to confine the competitions with the two-hop range for each node (line~\ref{ln:cvcomp}). The node that guards the most wins (line~\ref{ln:winner}); it hence changes its state (thus enter the second stage, line~\ref{ln:s1tos2}) and also notifies the guarded members to change their states (lines~\ref{ln:notecv} and \ref{ln:confirmcv}). As a result, these guarded members will not be counted in later competition (line~\ref{ln:count}). A node $v$ uses $v.\mathit{gd}$ to record its guarding node. For the guarded members, the algorithm has already terminated. At the end of S1, all nodes in \textbf{Connecting} state implicitly compose the \textbf{guardian set} $C$.

    A guarding node that enters the Steiner tree construction stage (S2) will periodically check whether other members are guarded and, for an already guarded one, get the shortest paths to its guarding node (lines~\ref{ln:pathqb} to \ref{ln:pathqe}). If the targeted multi-hop wireless network has a proactive routing protocol (e.g., OLSR \cite{OSLR}), then a node only needs to check its local routing table. Otherwise routing queries may need to be sent if the network uses a reactive routing protocol (e.g., AODV \cite{AODV} DSR or \cite{DSR}). The tree construction starts after all members are either guarded or enters S2. Our algorithm is similar to the Kruskal-based shortest path heuristic proposed in \cite{Bauer1996}. The basic idea is to construct a minimum spanning tree over the extended graph $G'(V, E')$, where an edge $(u,v) \in E'$ has a length of $\mathit{SP}(u \leftrightsquigarrow v)$ (the shortest path in $G$ between $u$ and $v$). As this heuristic is a distributed implementation of the 2-approximation algorithm proposed in \cite{Kou1981}, the resulting approximation ratio is indeed 2.

    The second stage (S2) starts with all the nodes in \textbf{Connecting} state, and it gradually finds a tree that spans these nodes. In order to eventually lead to a minimum spanning tree over $G'(V, E')$, only the shortest edge in $E'$ should be added each time. This can be easily distributed initially (when a node is not connected to any other node yet), as the local information on the shortest paths from a node is sufficient for it to decide which edge in $E'$ is to be added (line~\ref{ln:ssp}). However, after some nodes are connected to form several fragments, the local information is not sufficient anymore, hence we need a leader (the node in state \textbf{Merging}, chosen by the smallest $\mathit{id}$ principle in lines~\ref{ln:sid1} and \ref{ln:sid2}, and recorded by $v.\mathit{ld}$) and certain consensus for obtaining a common preferred node to connect to. This consensus is implemented in the procedure \textsc{Discover}$()$. Essentially, this procedure returns true if such a preferred node can be identified, then the leader (on behalf of the current fragment) sends a connection request to that node (line~\ref{ln:scane}). Otherwise the algorithm terminates, as the approximate Steiner tree has been constructed (line~\ref{ln:terminate}).\footnote{Actually, there are some details on pruning overlapping edges of different shortest paths, but we omit them for brevity.} It is important to note that a connection request is accepted only if both ends consider each other as preferred (lines~\ref{ln:ssps}, \ref{ln:scane}, and \ref{ln:match}); this is meant to avoid producing loops.

  \subsection{Performance Analysis} \label{sec:perf1}
    We examine the approximation ratio and the complexity of our algorithm in this section. In particular, we show that $T_A$ resulting from \textbf{Algorithm~\ref{alg:fixedplog}} is a $\left[4ln(\Delta+1)+7\right]$-approximation to $T_\mathit{opt}$, where $\Delta$ is the maximum node degree of $G$. Our result is an exponential improvement to the best known result \cite{Liang2009}. We also show that the complexity of \textbf{Algorithm~\ref{alg:fixedplog}} is polynomial in $|V|$.

    \subsubsection{Approximation Ratio} Our proof for the approximation ratio is presented in three steps. In \textit{\textbf{Lemma}~\ref{lma:inner}}, we provide an upper bound of the number of the internal nodes in $T_A$. In \textit{\textbf{Lemma}~\ref{lma:lowebound}}, we provide two lower bounds of the total energy consumption of any multicast tree for the MEAAM problem. These bounds are used by \textit{\textbf{Theorem}~\ref{thm:fixedratio}} to finally show the approximation ratio of \textbf{Algorithm~\ref{alg:fixedplog}}.
    %
    \begin{lemma}
      Let $T_I$ be the multicast tree spanning the nodes in $M$ such that $|\mathit{in}(T_I)|$ is minimized, we have: $|\mathit{in}({T_A})| \le \left[4\ln (\Delta+1)+6\right] \cdot |in({T_I})|$.
      \label{lma:inner}
    \end{lemma}
    \begin{IEEEproof}
      Let $T^*_S$ and $T_S$, respectively, be the minimum Steiner tree and the approximate Steiner tree (produced by S2 of \textbf{Algorithm~\ref{alg:fixedplog}}) that span the nodes in $C$. Given the fact that S2 of \textbf{Algorithm~\ref{alg:fixedplog}} is a distributed implementation of \cite{Kou1981}, $T_S$ is a 2-approximation of $T^*_S$, so $|\mathit{nd}(T_S)| \le 2|\mathit{nd}({T^*_S})|$. As $|\mathit{in}({T_A})| = |\mathit{nd}(T_S)|$ given the way $T_A$ is constructed, we have:
      \begin{eqnarray}
        |\mathit{in}({T_A})| \le 2|\mathit{nd}({T^*_S})|.
      \end{eqnarray}

      Let $T^-_I$ be the tree constructed from $T_I$ by pruning all the degree-one nodes in $T_I$ (we assume $|\mathit{nd}(T^-_I)| \neq 0$, otherwise, the proof becomes trivial). For any node $u \in C \backslash \mathit{nd}({T^-_I})$, we can find a node $u_1 \in M$ such that $u$ is adjacent to $u_1$. If $u_1$ is not in $T^-_I$, then there must exist a node $u_2$ in $T^-_I$ such that $u_2$ is adjacent to $u_1$. In other words, any node in $C \backslash \mathit{nd}({T^-_I})$ can be connected to $T^-_I$ by a path whose length is no more than 2. Therefore, we have:
      \begin{eqnarray}
        |\mathit{nd}({T^*_S})| \le |\mathit{nd}({T^-_I})| + 2|C| = |\mathit{in}({T_I})| + 2|C|.
      \end{eqnarray}

      For any node $v\in B$, we define $\sigma(v)=\mathit{nb}^+(v,G)\cap M$. Clearly, we have $|\sigma(v)|\leq \Delta+1$. Let $C^* \subseteq B$ be the guardian set in $G$ that contains the minimum number of nodes. Since S1 of \textbf{Algorithm~\ref{alg:fixedplog}} implements the greedy set covering algorithm proposed in \cite{Cormen2001}, we have:
           %
      %
      \begin{eqnarray}
        |C| &\le& \left[\ln \left(\max_{v \in B}|\sigma (v)|\right) + 1\right] \cdot |C^*| \nonumber \\
        &\le& \left[\ln (\Delta  + 1) + 1\right] \cdot |{C^*}|.
      \end{eqnarray}

      Note that $\mathit{nd}({T^-_I})\cap B$ is also a guardian set. so
      \begin{eqnarray}
        |{C^*}| \le |\mathit{nd}({T^-_I})| = |\mathit{in}({T_I})|.
      \end{eqnarray}

      Summarizing all the results obtained by far, we have:
      \begin{eqnarray}
        |\mathit{in}(T_A)| &\le& 2(|\mathit{in}({T_I})| + 2|C|) \nonumber \\
        &\le& 2|\mathit{in}({T_I})| + 4\left[\ln(\Delta +1) + 1\right] \cdot |\mathit{in}({T_I})| \nonumber \\
        &=& \left[4\ln (\Delta +1) + 6\right] \cdot |\mathit{in}({T_I})|, \nonumber
      \end{eqnarray}
      this bounds $|\mathit{in}({T_A})|$ from above by $|\mathit{in}({T_I})|$.
    \end{IEEEproof}

    \begin{lemma}
      For any multicast tree $T$ in $G$ spanning the nodes in $M$, we have:
      \begin{enumerate}
        \item $\Psi (T) \ge k \cdot \left[{\varepsilon _s} + (|M| - 1) \cdot {\varepsilon _r}\right]$
        \item $\Psi (T) \ge |\mathit{in}(T)| \cdot ({\varepsilon _s} + {\varepsilon _r}) \cdot k$
      \end{enumerate}
      \label{lma:lowebound}
    \end{lemma}
    \begin{IEEEproof}
      Each node $u \in M$ must transmit its own $p(u)$ data packets. Therefore, the total energy consumption for transmitting data packets in a group communication session is at least $\sum_{u \in M} {p(u) \cdot {\varepsilon_s}}=k \cdot {\varepsilon _s}$. Moreover, the total energy consumption for receiving the data packets originated from $u$ is $p(u) \cdot (|\mathit{nd}(T)| - 1) \cdot {\varepsilon _r}$. Therefore,
      \begin{eqnarray}
        \Psi (T) &\ge& k \cdot {\varepsilon _s} + \sum_{u \in M} {p(u) \cdot (|\mathit{nd}(T)| - 1) \cdot {\varepsilon _r}} \nonumber\\
        &\ge& k \cdot \left[{\varepsilon _s} + (|M| - 1) \cdot {\varepsilon _r}\right]. \nonumber
      \end{eqnarray}

      The total energy consumption for realizing a group communication session using $T$ can also be written as:
      \begin{eqnarray}
        \Psi (T) &=& \sum\nolimits_{u \in \mathit{lv}(T)} p(u) \cdot \left(|in(T)| + 1 \right) \cdot {\varepsilon _s} + \nonumber \\
        && \sum\nolimits_{u \in \mathit{in}(T) \cap M} {p(u) \cdot |\mathit{in}(T)|}  \cdot {\varepsilon _s} + \nonumber \\
        && k \cdot \left(|\mathit{nd}(T)| - 1\right) \cdot {\varepsilon _r} \nonumber \\
        &=& k \cdot |\mathit{in}(T)| \cdot \varepsilon _s + \sum\nolimits_{u \in \mathit{lv}(T)} p(u) \cdot \varepsilon _s + \nonumber \\
        && k \cdot (|\mathit{nd}(T)| - 1) \cdot {\varepsilon _r}. \label{eq:te}
      \end{eqnarray}

      Therefore:
      \begin{eqnarray}
        \Psi (T) &\ge& k \cdot |in(T)| \cdot {\varepsilon _s} + k \cdot \left(|in(T)|+|lv(T)| - 1\right) \cdot \varepsilon _r \nonumber \\
        &\ge& |in(T)| \cdot ({\varepsilon _s} + {\varepsilon _r}) \cdot k. \nonumber
      \end{eqnarray}
      These give the two claimed lower bounds on $\Psi(T)$.
    \end{IEEEproof}

    \begin{theorem}
      The multicast tree $T_A$ constructed by \textbf{Algorithm~\ref{alg:fixedplog}} has an approximation ratio of $4ln(\Delta+1)+7$ for the MEAAM problem.
      \label{thm:fixedratio}
    \end{theorem}
    \begin{IEEEproof}
      As $|\mathit{lv}(T_A)|\leq |M|$, using (\ref{eq:te}) we can get:
      \begin{eqnarray}
        \Psi(T_A) &=& k \cdot |\mathit{in}(T_A)| \cdot {\varepsilon _s} + \sum\nolimits_{u \in \mathit{lv}(T_A)} {p(u) \cdot } {\varepsilon _s} + \nonumber \\
        && k \cdot (|\mathit{nd}(T_A)| - 1) \cdot {\varepsilon _r} \nonumber \\
        &\le& k \cdot |\mathit{in}(T_A)| \cdot {\varepsilon _s} + \sum\nolimits_{u \in M} {p(u) \cdot } {\varepsilon _s} + \nonumber \\
        && k \cdot (|\mathit{in}(T_A)|+|M| - 1) \cdot {\varepsilon _r} \nonumber \\
        &\le& k \cdot |\mathit{in}(T_A)| \cdot ({\varepsilon _s} +{\varepsilon _r}) + \nonumber \\
        &&  k \cdot \left[\varepsilon _s + (|M| - 1) \cdot \varepsilon _r \right]. \label{eq:interm}
      \end{eqnarray}

      Let $\kappa=\Delta+1$. Using \textit{\textbf{Lemmas}~\ref{lma:inner}}, \textit{\ref{lma:lowebound}} and (\ref{eq:interm}), we have:
      \begin{eqnarray}
        \Psi(T_A) \!\!\!&\le&\!\!\! (4ln \kappa + 6) \cdot |\mathit{in}(T_I)| \cdot ({\varepsilon _s} +{\varepsilon _r}) \cdot k + \Psi(T_\mathit{opt}) \nonumber \\
        &\le&\!\!\! (4ln \kappa + 6) \cdot |\mathit{in}(T_\mathit{opt})| \cdot ({\varepsilon _s} +{\varepsilon _r}) \cdot k + \Psi(T_\mathit{opt}) \nonumber \\
        &\le&\!\!\! (4ln \kappa + 6) \cdot \Psi(T_\mathit{opt}) + \Psi(T_\mathit{opt}) \nonumber \\
        &=&\!\!\! \left[4\ln (\Delta+1) + 7\right] \cdot \Psi(T_\mathit{opt}), \nonumber
      \end{eqnarray}
      hence the claimed approximation ratio.
    \end{IEEEproof}

    \subsubsection{Algorithm Complexity} The first stage of \textbf{Algorithm~\ref{alg:fixedplog}} has a time complexity of $\mathcal{O}(\Delta)$, as for every $v: \mathit{nb}(v,G) \cap M \neq \emptyset$, at least one node in $\mathit{nb}(v,G)$ leaves $\mathbf{Inactive}$ state every round.\footnote{Our algorithm works in asynchronous settings; the concept of \textit{round} is used only for evaluating the time complexity.} In each round, all nodes in $\mathit{nb2}(v, G)$ sends a message, so the message complexity is $\mathcal{O}(\Delta \cdot |V|)$, where we use $|V|$ to bound from above the cardinality of $\mathit{nb2}(v, G)$. The second stage is a distributed Steiner tree algorithm. Its worst case time complexity and message complexity are $\mathcal{O}(\mathrm{D}(G) \cdot |M|)$ and $\mathcal{O}(|M| \cdot |V|)$, respectively, as (i) only one group member leaves the $\mathbf{Connecting}$ state in each round in the worst case (hence $|M|$ rounds in total), (ii) differing from the first stage where the length of each round is a constant, this time the length of a round is determined by the round-trip time of a path (hence bounded by $\mathrm{D}(G)$, the diameter of $G$), and (iii) the number of messages in each round is bounded by $|V|$.

  \subsection{Tree Maintenance}
    We discuss the tree maintenance with respect to the two individual stages separately. Actually, maintaining the shared tree simply for allowing the group communication session to continue is a separate issue that has been tackled in the literature (e.g., \cite{Chiang1998,Ji1998,Jia98}). Therefore, our focus is only on how to maintain the proven approximation ratio of the tree.

    When a member $v$ joins or leaves the group communication session, if it has no existing guarding node in $\mathit{nb}(v,G)$ (for join) or is the only guarded node (for leave), then the impact directly goes to the Steiner tree (which will be discussed later). Otherwise $v$ could directly join or leave in a localized manner by notifying its guarding node. The complication comes when a joining or leaving violates the greedy cover principle (thus affecting the approximation ratio). Therefore, we may need to re-execute S1 of \textbf{Algorithm~\ref{alg:fixedplog}} under those circumstances to maintain the greedy cover, hence maintain the proven approximation ratio.

    If a member joining or leaving does not lead to any changes in the existing guardian set $C$, then the approximate Steiner tree remains intact. If a member joining brings one more node to $C$, we may need to re-execute S2 of \textbf{Algorithm~\ref{alg:fixedplog}} in the worst cases (when this newly joined guarding node results in several short cuts). Fortunately, we also have another choice of having this node directly connected to a closest (in terms of shortest path) guarding node. According to \cite{ImaseW91}, this will lead to a $\mathcal{O}(\log|M|)$ approximation to Steiner tree, and in turn a $\mathcal{O}(\ln|M| \cdot \ln\Delta)$ approximation to MEAAM. If a member leaving removes a node from $C$, it may partition the tree into several fragments. The existing procedures of \textbf{Algorithm~\ref{alg:fixedplog}} can take care of this case as far as a new leader is elected for each fragment, as this is exactly the same as the fragment merging phase of the Steiner tree construction stage.

\section{MEGCOM-CFP: Constant Approximation for MEAAM with Fixed tx Power} \label{sec:conftxp}
  The algorithm proposed in Sec.~\ref{sec:logftxp} has a relatively high message complexity as we involve non-members in both stages. In this section, we propose a simplified algorithm that only involve group member in the first stage. Interestingly, we can show that this simplified algorithm has a constant approximation ratio.

  \subsection{The Algorithm}
    As this new algorithm share the same second stage (S2) with \textbf{Algorithm~\ref{alg:fixedplog}}, we only show the updated first stage (S1) in \textbf{Algorithm~\ref{alg:fixedpcon}} and then explain details accordingly.
    \begin{algorithm}[h!]
      \KwIn{$G=(V,E)$ and $M$. For each node $v \in V$, $v.\mathit{state} = \mathbf{Inactive}$;~~$v.\mathit{nb} = \mathit{nb}(v,G)$}
      \KwOut{An approximate multicast tree $T'_A$}
      \hrule
      \textbf{task}~\textsc{Scan}$()$  \hfill /* \textsf{\small executed periodically} */ \\
      \If {$v.\mathit{state} = \mathbf{Inactive} \wedge v \in M$} {
        \textsc{sendCompete}$(v.\mathit{id})$ to $\forall u \in v.\mathit{nb}$ \\
      }
      \hrule
      \textbf{upon}~\textsc{recvCompete}$(u.\mathit{id})$ \\
      \uIf {$v.\mathit{state} = \mathbf{Inactive}$} { \label{ln:loser}
        \lIf {$v.\mathit{id} > u.\mathit{id}$} {
          \textsc{sendAccept}$(v.\mathit{id})$ to $u$  \label{ln:acceptcv}
        }
      }
      \lElse {
        \textsc{sendACK}$(v.\mathit{id}, v.\mathit{state})$ \label{ln:announce} \\
      }
      \hrule
      \textbf{upon}~\textsc{recvAccept}$(u.\mathit{id})$ from $\forall u \in v.\mathit{nb}$ \label{ln:allaccept} \\
      $v.\mathit{state} \leftarrow \mathbf{Connecting};~v.\mathit{gd} \leftarrow v$ \label{ln:joinc}
      \hrule
      \textbf{upon}~\textsc{recvACK}$(u.\mathit{id}, u.\mathit{state})$ \\
      \uIf{$u.\mathit{state} = \mathbf{Connecting}$} {
        $v.\mathit{state} \leftarrow \mathbf{Treed};~~v.\mathit{gd} \leftarrow u$;~~mark the edge to $u$ \label{ln:covered}
      }
      \lElseIf {$u.\mathit{state} = \mathbf{Treed}$} {
        $v.\mathit{nb} \leftarrow v.\mathit{nb} \backslash \{u\}$ \label{ln:updatenb}
      }
      \caption{Finding an approximate multicast tree $T'_A$ for the MEAAM problem} \label{alg:fixedpcon}
    \end{algorithm}

    The objective is again to identify a guardian set $C'$ for $M$, but the algorithm differs from \textbf{Algorithm~\ref{alg:fixedplog}} in that $C'$ is construct from $M$, i.e., $C' \subseteq M$. Similar to \textbf{Algorithm~\ref{alg:fixedplog}}, all the nodes that eventually change to $\mathbf{Connecting}$ state are in $C'$. The competition is again based on the smallest $\mathit{id}$ principle. A node in $\mathbf{Inactive}$ state accepts any neighbor to be its guardian if its own $\mathit{id}$ is larger (lines~\ref{ln:loser} to \ref{ln:acceptcv}). If a node receives the acceptance from all its $\mathbf{Inactive}$ neighbors, it joins the guardian set $C'$ (lines~\ref{ln:allaccept} to \ref{ln:joinc}) by changing its state to $\mathbf{Connecting}$. A node that is not in $\mathbf{Inactive}$ state always acknowledge the competition message by announcing its current state (line~\ref{ln:announce}). A node either becomes guarded if it receives an acknowledgement that carries a $\mathbf{Connecting}$ state (line~\ref{ln:covered}) or updates its inactive neighbor table if the state carried in the acknowledgement is $\mathbf{Treed}$. Note that this algorithm only uses one-hop communications and finds an arbitrary guardian set (sufficient to achieve a constant approximation ratio), but we could also apply a similar procedure in \textbf{Algorithm~\ref{alg:fixedplog}} to find a greedy guardian set, which needs two-hop communications.

  \subsection{Performance Analysis}
    The approximation ratio of $T'_A$ obtained by \textbf{Algorithm~\ref{alg:fixedpcon}} is immediate from the following theorem.
    \begin{theorem}
      $\Psi(T'_A)\leq 13\Psi(T_\mathit{opt})$
      \label{thm:distributed}
    \end{theorem}
    \begin{IEEEproof}
      Let $T_I$ again be the multicast tree spanning the nodes in $M$ such that $|\mathit{in}(T_I)|$ is minimized. The outcome of \textbf{Algorithm~\ref{alg:fixedpcon}} implies that the nodes in $C'$ are mutually independent. Therefore, any node in $\mathit{in}(T_I)$ can be adjacent to at most 5 nodes in $C'$ \cite{Wan2002}. Moreover, as any node $u \in C'$ is a member of $M$, $u$ is either in $\mathit{in}(T_I)$ or is adjacent to certain node in $\mathit{in}(T_I)$. So we can get:
      \begin{eqnarray}
        |C'|\le 5|\mathit{in}(T_I)| \label{eq:5t}
      \end{eqnarray}

      Let the minimum Steiner tree and approximate Steiner tree (produced by S2 of \textbf{Algorithm~\ref{alg:fixedpcon}}) spanning the nodes in $C'$ be $\acute{T}^*_S$ and $\acute{T}_S$, respectively. We have:
      \begin{eqnarray}
        |\mathit{in}(T'_A)| = |\mathit{nd}(\acute{T}_S)| \le 2|\mathit{nd}(\acute{T}^*_S)|
      \end{eqnarray}

      Since any nodes in $C'$ is either in $\mathit{in}(T_I)$ or is adjacent to certain node in $in(T_I)$, we can find a Steiner tree spanning the nodes in $C'$ whose number of nodes is at most $|C'|+|\mathit{in}(T_I)|$. Therefore, we have:
      \begin{eqnarray}
        |\mathit{nd}(\acute{T}^*_S)|\le |C'|+|\mathit{in}(T_I)| \label{eq:sbd}
      \end{eqnarray}

      Combining (\ref{eq:5t})--(\ref{eq:sbd}), we can get:
      \begin{eqnarray}
        |\mathit{in}(T'_A)|\le 2(|C'|+|\mathit{in}(T_I)|) \le 12|\mathit{in}(T_I)|
      \end{eqnarray}

      The result by far gives an upper bound for $|\mathit{in}(T'_A)|$. Now we can apply \textit{\textbf{Lemma}~\ref{lma:lowebound}} and (\ref{eq:interm}) (similar to the proof of \textbf{Theorem~\ref{thm:fixedratio}}) to obtain $\Psi(T'_A)\leq 13\Psi(T_\mathit{opt})$.
    \end{IEEEproof}

    We omit the complexity analysis for this algorithm, as it is very similar to \textbf{Algorithm~\ref{alg:fixedplog}}, except that the constant length of a round in the first stage becomes shorter, as only one-hop communications are required. The tree maintenance becomes simpler, as the guardian set is not constructed in a greedy manner, the impact of member joining or leaving directly goes to the Steiner tree.

\section{MEGCOM-CAP: Constant Approximation for MEAAM with Adjustable tx Power} \label{sec:conatxp}
  When the tx power of each node is adjustable, the idea of guardian set does not work anymore, as whether a node can be guarded by another is not known before an actual tx power is chosen. Even if we have a shared multicast tree for the group communication session, any node $u$ in a multicast tree $T$ may need to use the tx power $\max\{d_{(u,v)}^\alpha |v \in \mathit{nb}(u,T)\}$ to transmit the data packets. Therefore, different nodes may use different tx power to forward the data packets. This makes the MEAAM problem more complicated.

  To design an approximation algorithm for the MEAAM problem in the adjustable-tx-power case, we use a constructive proof to show the existence of a constant approximation algorithm, which naturally suggests one possible algorithm to construct the multicast tree. We still try to build a multicast tree whose sum of tx power of the internal nodes is the minimized, because the internal nodes still have a heavier forwarding load than the degree-one nodes for any multicase tree spanning $M$. Let $\lambda(u,T) = \max\{d_{(u,v)}^\alpha |v \in \mathit{nb}(u,T)\}$, and let $T_W$ be a multicast tree such that $\Theta ({T_W}) = \sum_{u \in \mathit{in}({T_W})} {\lambda (u,{T_W})}$ is minimized. Our idea is to find a multicast tree that approximates $T_W$ and in turn approximates $T_\mathit{opt}$. The trick is to identify a quantitative relation between $\Theta$ and $\Psi$. This is done by \textit{\textbf{Theorem}~\ref{thm:relation}}, which in turn rely on the upper bounds of the total energy consumption for transmitting and receiving data, respectively, derived in \textit{\textbf{Lemma}~\ref{lma:upbdtrans}} and \textit{\textbf{Lemma}~\ref{lma:upbdrecv}}.

  \begin{lemma}
    For any multicast tree $T$ spanning the nodes in $M$, the total energy consumption of transmitting all data packets using $T$ in a group communication session is at most $2k \cdot \Theta(T)$.
    \label{lma:upbdtrans}
  \end{lemma}
  \begin{IEEEproof}
    Each internal node in $\mathit{in}(T)$ must transmit every data packet sent by each member node. So the total energy consumption for transmitting data packets by the internal nodes in $T$ is:
    \begin{eqnarray}
      \sum\nolimits_{u \in in(T)} {\lambda (u,T) \cdot \sum\nolimits_{v \in M} {p(v)} } = k\cdot \Theta(T). \label{eq:bdtin}
    \end{eqnarray}

    Any degree-one node in $T$ has a unique neighboring node in $T$. Therefore, there exists a function $f:\mathit{lv}(T)\rightarrow \mathit{in}(T)$ such that $f(u)$ is the unique internal node adjacent to $u: \forall u \in \mathit{lv}(T)$. Let $Z = \{ f(u)|u \in \mathit{lv}(T)\}$. Clearly, for any degree one node $u \in T$, we have $\lambda(u,T) \le \lambda(f(u),T)$. The total energy consumption for transmitting data packets by the degree-one nodes in $T$ is:
    \begin{eqnarray}
      &&\sum\nolimits_{u \in \mathit{lv}(T)}{\lambda (u,T) \cdot p(u)} \nonumber \\
      &=& \sum\nolimits_{v \in Z} \sum\nolimits_{u \in \{w|w\in \mathit{lv}(T),f(w) = v\}} \lambda (u,T) \cdot p(u) \nonumber \\
      &\le& \sum\nolimits_{v \in Z} {\sum\nolimits_{u \in \mathit{lv}(T)} {\lambda (v,T) \cdot p(u)} } \nonumber \\
      &\le& k \cdot \sum\nolimits_{v \in Z} {\lambda (v,T)} \nonumber \\
      &\le& k \cdot \Theta (T). \label{eq:bdtlv}
    \end{eqnarray}

    Combining (\ref{eq:bdtin}) and (\ref{eq:bdtlv}), the lemma follows.
  \end{IEEEproof}

  \begin{lemma}
    For any multicast tree $T$ spanning the nodes in $M$, the total energy consumption of receiving all data packets using $T$ in a group communication session is at most $\Psi ({T_\mathit{opt}}) + k \cdot \Theta(T)$.
    \label{lma:upbdrecv}
  \end{lemma}
  \begin{IEEEproof}
    Based on our assumption that $\forall v \in T$, $\varepsilon _r \le \lambda(v,T)$, we can get:
    \begin{eqnarray}
      |\mathit{in}(T)| \cdot k \cdot {\varepsilon _r} &=& k \cdot \sum\nolimits_{v \in \mathit{in}(T)} {{\varepsilon _r}} \nonumber \\
      &\le& k \cdot \sum\nolimits_{v \in \mathit{in}(T)} \lambda (v,T) \nonumber \\
      &=& k \cdot \Theta (T). \nonumber
    \end{eqnarray}

    The total energy consumption for receiving all the data packets by the nodes in $T$ is $k \cdot (|\mathit{nd}(T)| - 1) \cdot \varepsilon _r$, and as $|\mathit{lv}(T)|\le |M| \le |\mathit{nd}(T_\mathit{opt})|$, we have:
    \begin{eqnarray}
      && k \cdot (|\mathit{nd}(T)| - 1) \cdot \varepsilon _r \nonumber \\
      &=& k \cdot \left[|\mathit{lv}(T)| + |\mathit{in}(T)| - 1\right] \cdot {\varepsilon _r} \nonumber \\
      &\le& k \cdot \left[|\mathit{nd}({T_\mathit{opt}})| - 1\right] \cdot \varepsilon _r + k \cdot |\mathit{in}(T)|\cdot \varepsilon _r \nonumber \\
      &\le& \Psi ({T_\mathit{opt}}) + k \cdot \Theta(T),
    \end{eqnarray}
    hence the claimed upper bound follows.
  \end{IEEEproof}

  \begin{theorem}
    Let $T_W$ be a multicast tree spanning the nodes in $M$ such that $\Theta(T_W)$ is minimized. For any multicast tree $T$, if $\Theta(T)\le \rho \cdot \Theta(T_W)$, then $\Psi (T)\le (3\rho +1)\cdot \Psi(T_\mathit{opt})$
    \label{thm:relation}
  \end{theorem}
  \begin{IEEEproof}
    With \textit{\textbf{Lemma}~\ref{lma:upbdtrans}} and \textit{\textbf{Lemma}~\ref{lma:upbdrecv}}, we have:
    \begin{eqnarray}
      \Psi (T) &\le& 3k \cdot \Theta (T) + \Psi({T_\mathit{opt}}) \nonumber \\
      &\le& 3\rho k \cdot \Theta ({T_W}) + \Psi({T_\mathit{opt}}) \nonumber \\
      &\le& 3\rho k \cdot \Theta ({T_\mathit{opt}}) + \Psi(\mathit{T_{opt}}) \nonumber \\
      &\le& (3\rho + 1) \Psi(\mathit{T_{opt}}),
    \end{eqnarray}
    hence the theorem follows.
  \end{IEEEproof}

  \vspace{1ex}\noindent\textit{Remark: This theorem is itself very important. It shows that, if we could find a tree $T$ that approximates $T_W$ within a constant ratio, $T$ is also a constant approximation to $T_{opt}$.}\vspace{1ex}

  In this paper, we only suggest a straightforward algorithm that achieves a constant approximation ratio. In fact, we will show that, if we apply the distributed Steiner tree algorithm given in S2 of \textbf{Algorithm~\ref{alg:fixedplog}} to span $M$ directly, the resulting tree $\ddot{T}_S$ approximates $T_W$ within a constant ratio.
  \begin{lemma}
    Let $\ddot{T}_S$ be a 2-approximation Steiner tree in $G$ spanning $M$, we have: $\Theta (\ddot{T}_S) \le 48\Theta (T_W)$.
    \label{lma:steiner}
  \end{lemma}
  \begin{IEEEproof}
    For any multicast tree $T$ spanning the nodes in $M$, we denote by $\zeta(T)$ the sum of the weights (tx power) of the edges in $T$. For any node $u$ in $\mathit{in}(T_W)$, we denote by $T_W^{(u)}$ the Euclidean minimum spanning tree of the nodes in $\{u\} \cup \mathit{nb}(u,{T_W})$. According to \cite{Wan2004}, we have:
    \begin{eqnarray}
      \zeta\left(T_W^{(u)}\right) \le c \cdot \lambda (u,{T_W}),~~6\le c\le 12.
    \end{eqnarray}

    Let $G'$ be the graph constructed by superposing the $T_W^{(u)}$'s for all $u\in \mathit{in}(T_W)$. Let $T_z$ be an arbitrary spanning tree of $G'$. We can get:
    \begin{eqnarray}
      \zeta ({T_z}) &\le& \sum\nolimits_{u \in \mathit{in}({T_W})} {\zeta \left(T_W^{(u)}\right)} \nonumber \\
      &\le& c \cdot \sum\nolimits_{u \in \mathit{in}({T_W})} {\lambda (u,{T_W})} \nonumber \\
      &=& c \cdot \Theta ({T_W}).
    \end{eqnarray}

    Let $\ddot{T}_S^*$ be the minimum Steiner tree in $G$ spanning the nodes in $M$. We have:
    \begin{eqnarray}
      \zeta ({\ddot{T}_S}) \le 2\zeta (\ddot{T}_S^*) \le 2\zeta ({T_z}).
    \end{eqnarray}

    Note that the weight of any edge in $\ddot{T}_S$ can be counted at most twice in $\Theta ({\ddot{T}_S})$, so
    \begin{eqnarray}
      \Theta ({\ddot{T}_S}) \le 2\zeta ({\ddot{T}_S}).
    \end{eqnarray}

    The claimed approximation ratio follows by combining all these inequalities.
  \end{IEEEproof}

  Finally, using Theorem~\ref{thm:relation} and Lemma~\ref{lma:steiner} we can easily get:
  \begin{theorem}
    Our distributed Steiner tree algorithm on $M$ approximates the MEAAM problem under the adjustable tx power case with a constant approximation ratio of 145. 
    \label{thm:adjustable}
  \end{theorem}

\section{Simulations} \label{sec:sim}
  In this section, we present the simulation results of our algorithms. We implement a simulator using C++ and use it to study how the performance of different MEGCOM algorithms is affected by various network parameters such as the network size, the node density and the percentage of group members. In the simulations, the network nodes are randomly deployed in a square area according to a certain node density, where the node density is defined as 1 if $|V|$ nodes are deployed upon a $\sqrt{|V|}\times\sqrt{|V|}$ square. The group members are selected from $V$ following a Bernoulli distribution, and the number of packets originated from each group member is chosen from a uniform distribution $\mathcal{U}(1, 100)$. For each setting of the network parameters, we generate 100 network instances and show the mean values of the simulation results. Since our MEGCOM-CAP algorithm for the adjustable transmission power case is the approximate Steiner tree algorithm, which has already been shown in experiments to be superior to other algorithms \cite{Liang2009}, we only study the performance of MEGCOM-LFP and MEGCOM-CFP under the fixed transmission power case with $\varepsilon_s=200$ and $\varepsilon_r=20$.

  For comparison, we build different multicast trees for the MEAAM problem in the simulations, using MEGCOM-LFP (\textbf{Algorithm~\ref{alg:fixedplog}}), MEGCOM-CFP (\textbf{Algorithm~\ref{alg:fixedpcon}}), the Shortest Path Tree (SPT) algorithm, and the Steiner Tree algorithm. In the SPT algorithm, a root node is randomly selected from the group members, and the shared multicast tree is construct by aggregating the shortest paths from the root node to other group members. In the implementation of the Steiner Tree algorithm, we adopt the approximation algorithm proposed by \cite{Kou1981}; this is the Steiner tree algorithm also used in \cite{Liang2009} as the solution to the MEAAM problem, and used in \cite{Wan2004} as the solution to the min-energy one-to-many multicasting problem. We refrain from implementing other multicasting algorithms for the MEAAM problem such as the algorithms proposed by \cite{Chiang1998,Ji1998}, since they have been shown to be inferior to the Steiner Tree algorithm \cite{Liang2009}.

\begin{figure*}[t]
      \begin{center}
        \includegraphics[width=.325\textwidth]{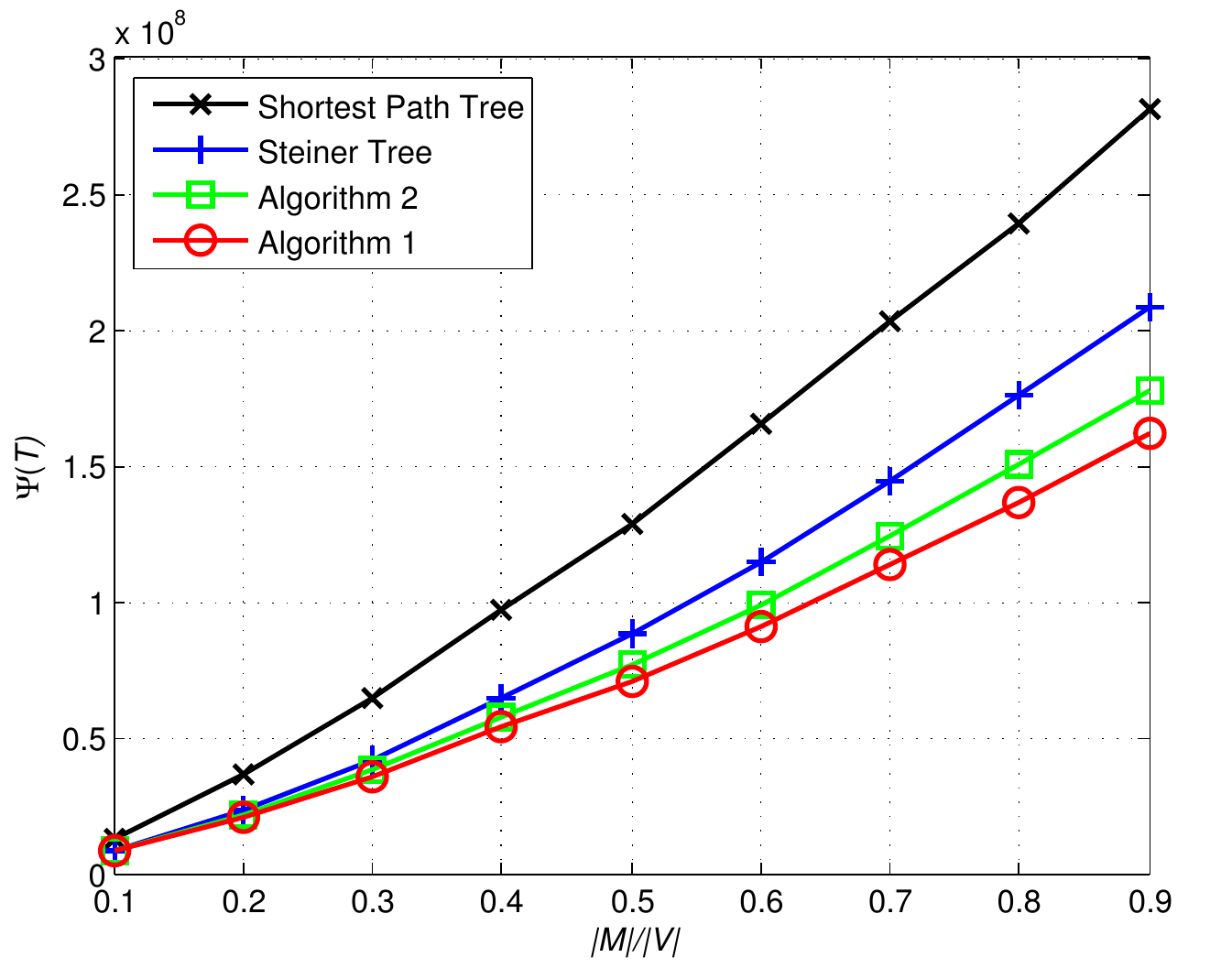}
	    \includegraphics[width=.325\textwidth]{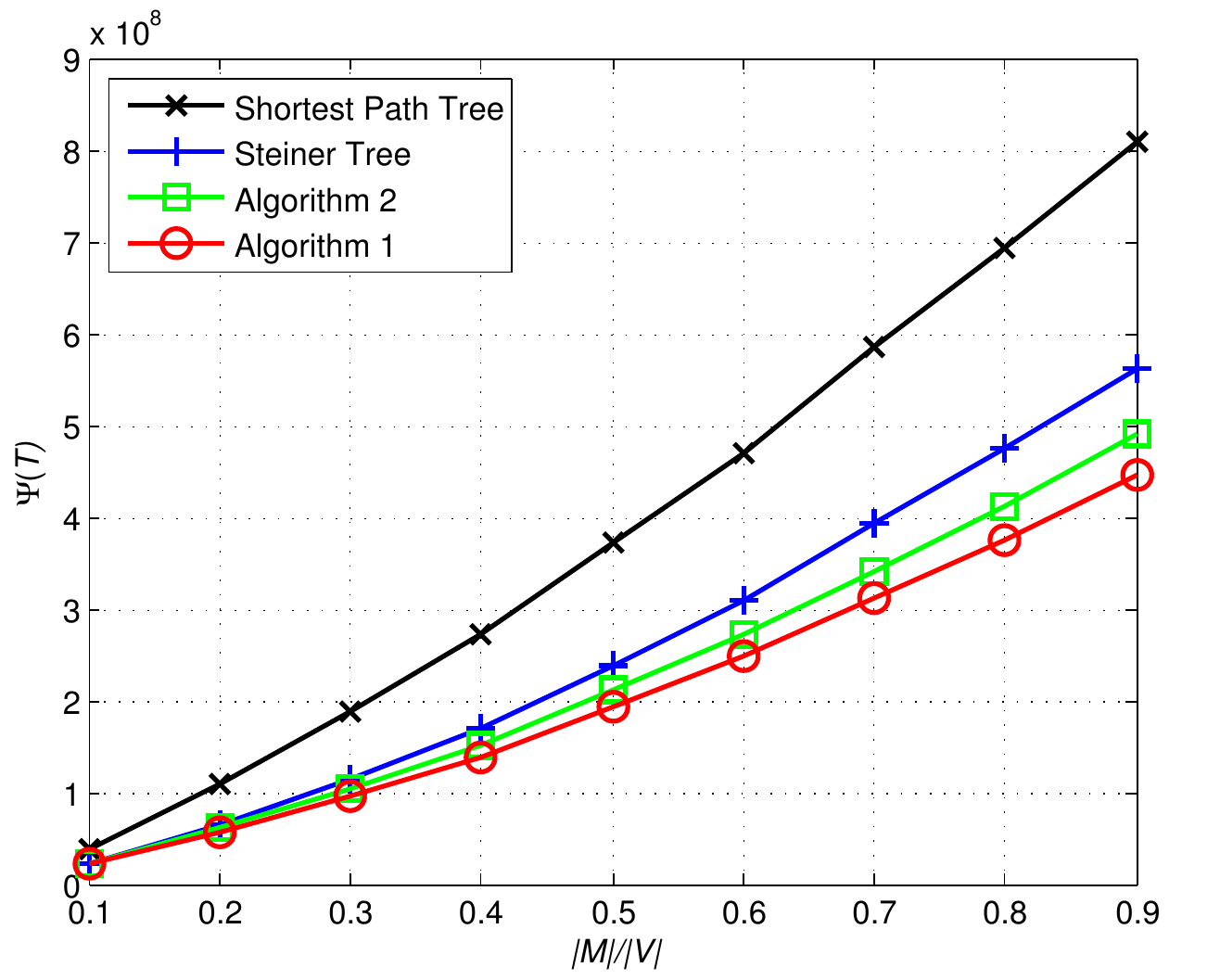}
        \includegraphics[width=.325\textwidth]{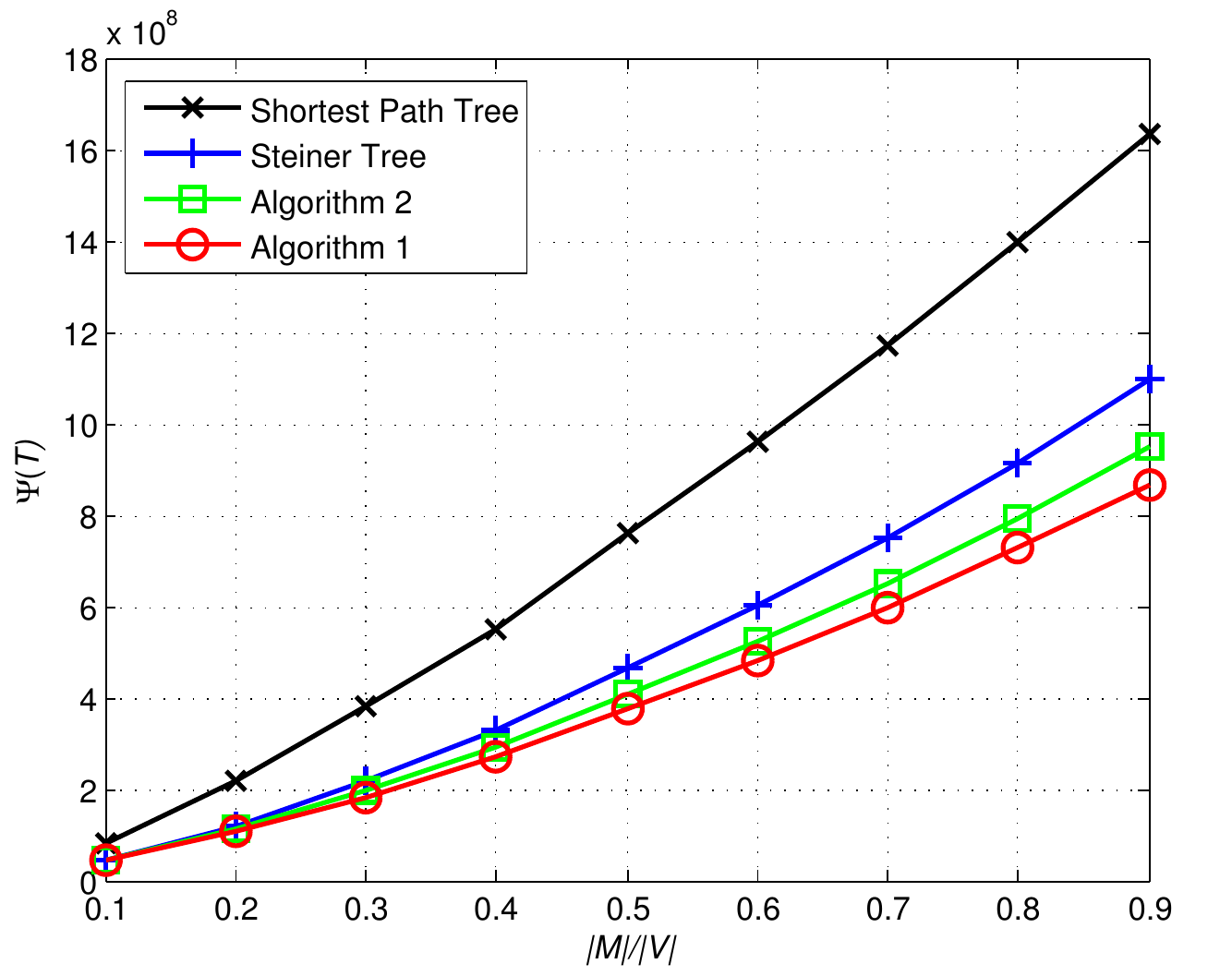}
        \parbox{.325\textwidth}{\center\scriptsize~(a) 300-node networks}
        \parbox{.325\textwidth}{\center\scriptsize~(b) 500-node networks}
        \parbox{.325\textwidth}{\center\scriptsize~(c) 700-node networks}
        \caption{Comparing different algorithms for building multicast trees with the network density fixed to be 2.}
        \label{fig:ratio}
      \end{center}
      \vspace{-.9ex}
    \end{figure*}

  In Figure~\ref{fig:ratio}, the network density is fixed to 1 and the transmission range of each node is set to 2 to maintain the network connectivity. The network size is set to 300, 500, and 700 in Figure~\ref{fig:ratio}(a)-(c), respectively, and we vary the percentage of the group members from 10\% to 90\%.
    \begin{figure*}[t]
      \begin{center}
        \includegraphics[width=.325\textwidth]{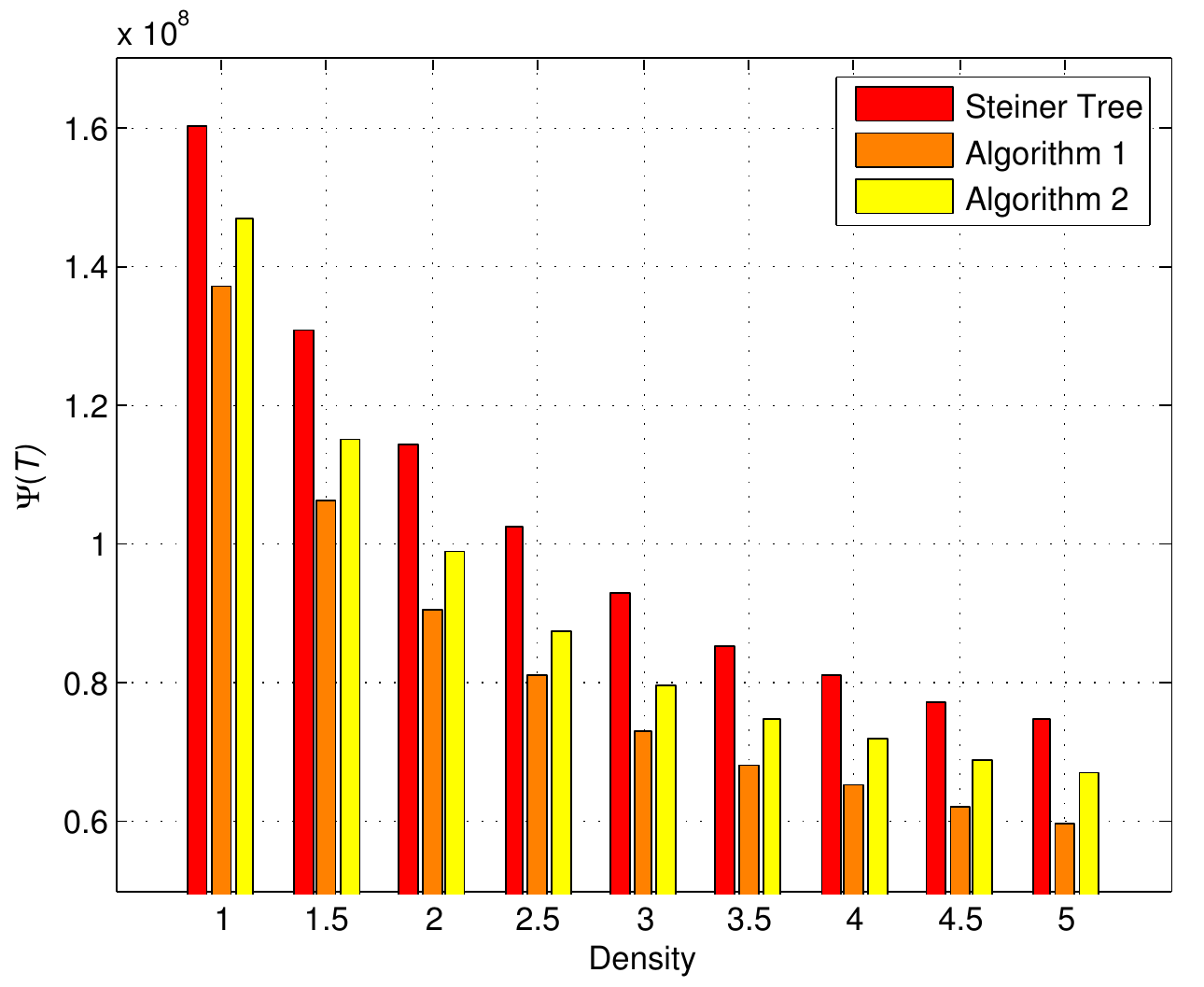}
	    \includegraphics[width=.325\textwidth]{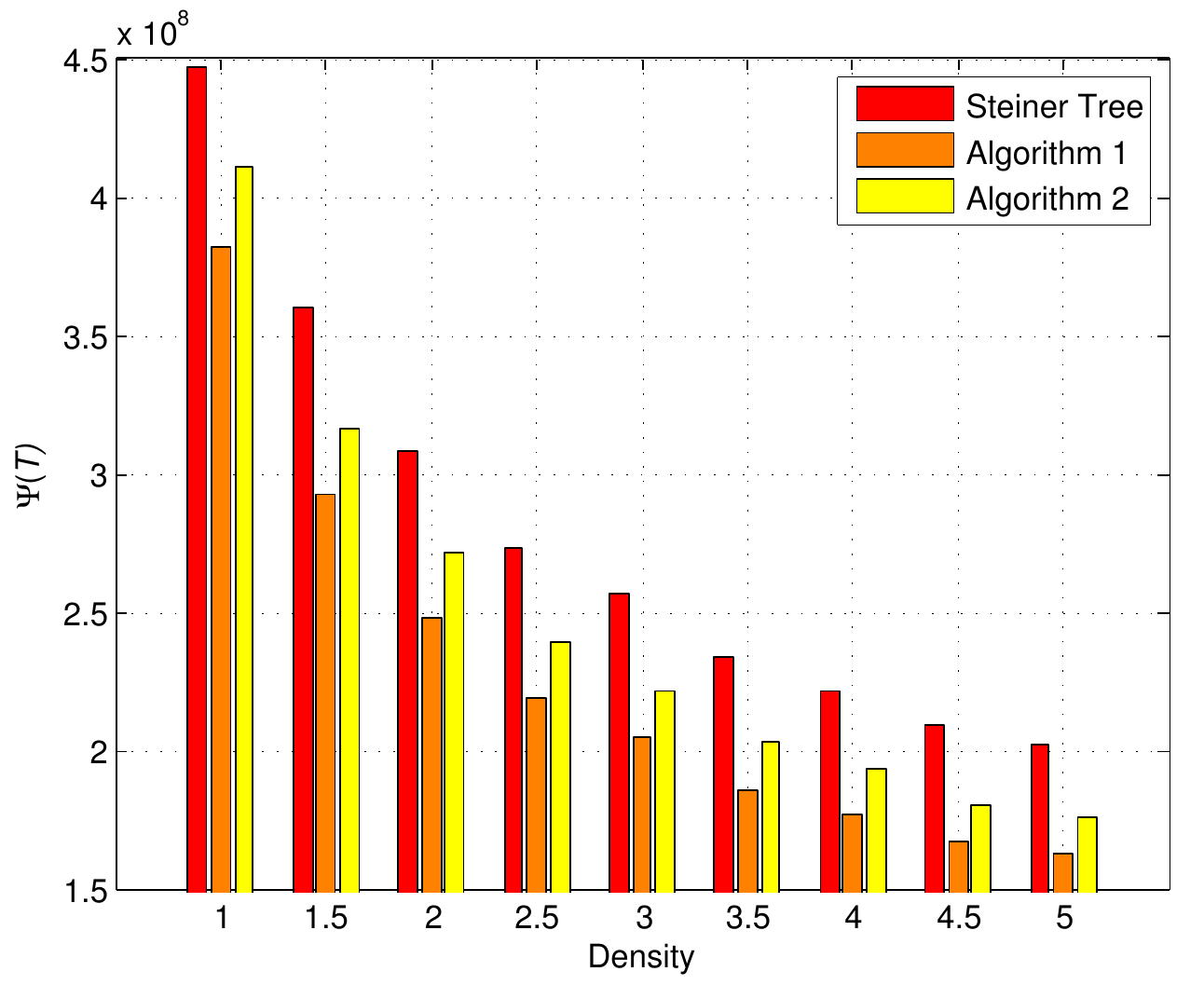}
        \includegraphics[width=.325\textwidth]{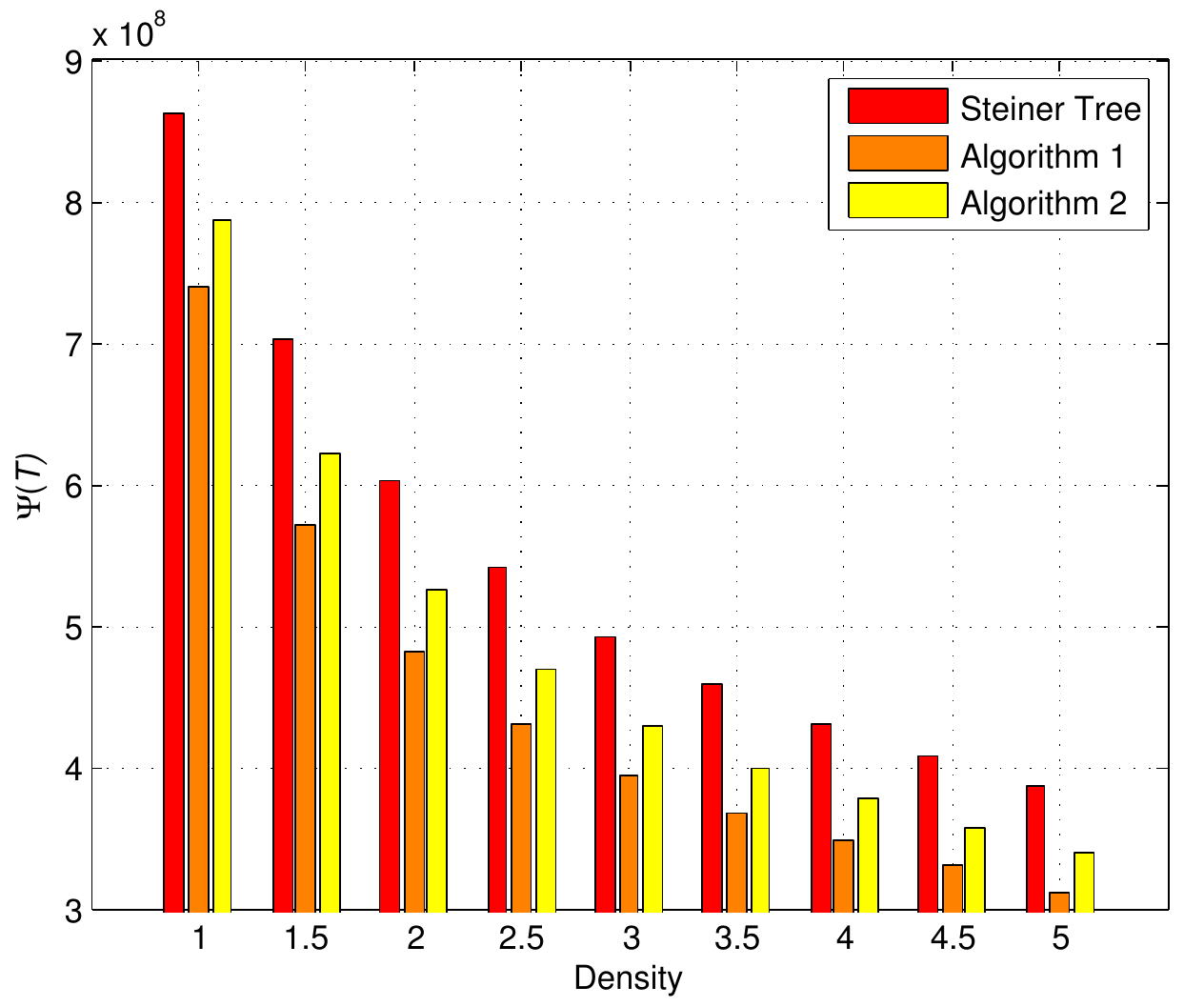}
        \parbox{.325\textwidth}{\center\scriptsize~(a) 300-node networks}
        \parbox{.325\textwidth}{\center\scriptsize~(b) 500-node networks}
        \parbox{.325\textwidth}{\center\scriptsize~(c) 700-node networks}
        \caption{Comparing different algorithms for building multicast trees with the percentage of group members fixed to be 60\%.}
        \label{fig:density}
      \end{center}
      \vspace{-.9ex}
    \end{figure*}
  It can be seen from Figure~\ref{fig:ratio} that the SPT algorithm always consumes the highest energy. This can be explained by the fact that the SPT tree is composed of shortest paths from one group member to the rest, introducing a large number of internal nodes that are not group members. Moreover, we can also see that \textbf{Algorithm~\ref{alg:fixedplog}} and \textbf{Algorithm~\ref{alg:fixedpcon}} both outperform the Steiner tree algorithm. This phenomenon validates our analysis on approximation radio in Section~\ref{sec:logftxp} and Section~\ref{sec:conftxp}. The main reason for this phenomenon is that \textbf{Algorithm~\ref{alg:fixedplog}} and \textbf{Algorithm~\ref{alg:fixedpcon}} both reduce the internal nodes of the shared multicast tree, whose energy consumption is the predominant part of the total energy consumption of a group communication session. Finally, we observe from Figure~\ref{fig:ratio} that our algorithms perform better when the percentage of group members increases. Actually, when 90\% network nodes become group members, \textbf{Algorithm~\ref{alg:fixedplog}} saves up to 40\% energy cost compared with the SPT algorithm, or 25\% compared with the Steiner tree algorithm. This can be understood by the fact that more energy are conserved by our algorithms when more data packets are involved in the group communication session.

  In Figure~\ref{fig:density}, we study the impact of node density on the performance of different algorithms. This time we remove the results for the SPT algorithm, as it perform worst in the previous comparisons. The percentage of group members is fixed to 60\%, and the node density scales from 1 to 5 with an increment of 0.5. Again, the network size is set to 300, 500, and 700 in Figure~\ref{fig:density}(a)-(c), respectively. We observe that \textbf{Algorithm~\ref{alg:fixedplog}} and \textbf{Algorithm~\ref{alg:fixedpcon}} always perform better than the Steiner tree algorithm in Figure~\ref{fig:density}. In particular, \textbf{Algorithm~\ref{alg:fixedplog}} saves 15\%$\sim$20\% energy compared with the Steiner tree algorithm. This demonstrates that the superiority of our MEGCOM algorithms persists when the node density changes. Here we use density as a rough indicator of $\Delta$, as generating degree constrained networks is non-trivial. And even if we generated such graphs, they could hardly reflect any practical WSN deployments.

  An interesting fact revealed by the simulations is that \textbf{Algorithm~\ref{alg:fixedplog}} always performs better than \textbf{Algorithm~\ref{alg:fixedpcon}}, although the latter has a better theoretical approximation ratio than the former (see \textit{\textbf{Theorem}~\ref{thm:fixedratio}} and \textit{\textbf{Theorem}~\ref{thm:distributed}}).\footnote{One generally considers a constant approximation ratio to be better than a logarithmic ratio. However, as an approximation ratio only serves as a metric to measure the worst case performance of an algorithm, the average performance of the algorithm in practice may not be well characterized by its approximation ratio.} We can attribute this effect to the fact that the guardian set is searched from a larger solution space in \textbf{Algorithm~\ref{alg:fixedplog}} than in \textbf{Algorithm~\ref{alg:fixedpcon}}, which should on average result in less internal nodes in the shared multicast tree constructed by \textbf{Algorithm~\ref{alg:fixedplog}}, hence a lower total energy consumption of a group communication session using the tree and a better performance of \textbf{Algorithm~\ref{alg:fixedplog}} in average sense.


\section{Conclusion} \label{sec:con}
  In this paper, we have studied the Minimum-Energy All-to-All Multicasting (MEAAM) problem in multi-hop wireless networks, where the transmission power of each wireless node could be either fixed or adjustable. Since the MEAAM problem is NP-complete, we have provided a set of distributed approximation algorithms under our MEGCOM (Min-Energy Group COMmunication) framework. For each algorithm, we have proven its approximation ratio with respect to the optimal solution. Our approximation ratios are significantly better than those of the best-known approximation algorithms for the MEAAM problem. We have further performed extensive simulations to validate our theoretical analysis and also to confirm the energy efficiency of our MEGCOM algorithms.



\bibliographystyle{IEEEtran}
\bibliography{IEEEabrv,Mylib}

\end{document}